\documentclass[aps,12pt,final,notitlepage,oneside,nobibnotes,nofootinbib,superscriptaddress,onecolumn,noshowpacs,amsmath,amssymb]{revtex4}

\usepackage[english]{babel}
\usepackage[utf8x]{inputenc}
\usepackage[T1]{fontenc}

\usepackage{amsmath}
\usepackage{graphicx}
\usepackage[colorinlistoftodos]{todonotes}
\usepackage[colorlinks=true, allcolors=blue]{hyperref}
\usepackage{url}

\newcommand{\nder}{\!\!\!\not\partial}

\begin{document}

\title{Zero Points of Chiral Condensate}
\author{V.G. Ksenzov}
\affiliation{State Scientific Center Institute for Theoretical and Experimental Physics, Moscow, 117218, Russia}
\author{A.I. Romanov}
 \email{einzehl@gmail.com}

\affiliation{National Research Nuclear University, MEPHI, Moscow, 115409, Russia}

\begin{abstract}
We investigate a model with a massless fermion and a massive scalar field with the Yukawa interaction between these two fields. The model possess a discrete symmetry. The chiral condensate is calculated in one-loop approximation in $(1+1)$-dimensional spacetime. It was shown that the chiral condensate vanishes at two different values of the coupling constant. At the first of them there is a phase transition in which the discrete symmetry is restored. At the second zero of the chiral condensate it takes place a phase transition which distinguishes the positive $\mu^2$ from the negative $\mu^2$, where $\mu^2$ is an effective mass of the scalar field.
\end{abstract}

\maketitle

\section{Introduction}

It is fairly well-known that a dynamical chiral symmetry breaking is characterized by the chiral condensate which develops a non-vanish vacuum expectation value \cite{GL}, \cite{NL}, \cite{GN}. So we believe that the vanishing of the chiral condensate is associate with the restoration of symmetry, but this is not entirely true. It turns out that the chiral condensate may vanish at several points.

Really, in our previous papers \cite{KR1}, \cite{KR2} we investigated a model of a self-interacting massive scalar and a massless fermion fields with the Yukawa interaction in a $(1+1)$-dimensional spacetime. The lagrangian of the model is invariant under the discrete transformation. The chiral condensate is an order parameter to describe a phase transition related to a dynamical discrete symmetry breaking. We also found that the chiral condensate vanishes at some other point of the coupling constant is not associated with the discrete symmetry \cite{KR2}.

The purpose of this paper is to research what means the second zero of the chiral condensate. It is argued that in this case it takes place a phase transition which distinguishes the positive $\mu^2$ from the negative $\mu^2$, where $\mu^2$ is an effective mass of the scalar field. The phase transition of such a kind was discussed in \cite{W}, \cite{KR3}.

\section{Chiral condensate as an order parameter for two different phase transitions}

To realize what means vanishing of the chiral condensate at the second point of the coupling constant we analyze a specific field-theoretic model with the Lagrangian density given by
\begin{equation}\label{eq1}
  L=L_b+L_f=\frac{1}{2}(\partial_{\mu}\phi)^2-\frac{1}{2}m^2\phi^2(x)+i\bar{\psi}^a\nder\psi^a-g\phi(x)\bar{\psi}^a\psi^a,
\end{equation}
where $\phi(x)$ is a real scalar field, $\psi^a(x)$ is a massless fermion field and index $a$ runs from 1 to $N$. This model allows us to obtain an exact result in an analytical form.

The Lagrangian (\ref{eq1}) is invariant under the discrete transformation
\begin{equation}\label{eq2}
  \psi^a\to\gamma_5\psi^a,\;\bar{\psi}^a\to-\bar{\psi}^a\gamma_5,\;\phi\to-\phi.
\end{equation}
The symmetry (\ref{eq2}) is broken either by the chiral condensate or by the scalar field which develops a non-vanishing vacuum expectation value (v.e.v). As it was shown in \cite{GN} -- \cite{KR2}, in this model the chiral condensate is connected with v.e.v. of scalar field. It should be noted, that the symmetry (\ref{eq2}) takes place only at the classical level but quantum contributions violate it for any $g$, except $g=0$. In latter case the chiral condensate must vanishes. As we discussed in \cite{KR2}, the chiral condensate also vanishes at some other coupling constant is not associated with the discrete symmetry. To study this phenomenon we formally define the chiral condensate using the functional integral
\begin{equation}\label{eq3}
  \left\langle0|g\bar{\psi}^a\psi^a|0\right\rangle=\frac{1}{Z}\int D\phi D\bar{\psi}^aD\psi^ag\bar{\psi}^a\psi^a \exp\left(i\int d^2xL(x)\right),
\end{equation}
where $Z$ is a normalization constant. Eq.~(\ref{eq3}) is rewritten as
\begin{equation}\label{eq4}
  \left\langle0|g\bar{\psi}^a\psi^a|0\right\rangle=\frac{1}{Z}\int D\phi e^{i\int d^2xL_b(x)}i\frac{\delta}{\delta\phi}\int D\bar{\psi}^aD\psi^a \exp\left(i\int d^2xL_f(x)\right).
\end{equation}
The fermionic Lagrangian is quadratic in the field and we can integrate over them, getting
$$\left\langle0|g\bar{\psi}^a\psi^a|0\right\rangle=\frac{1}{Z}\int D\phi \frac{g^2 N\phi}{2\pi}\ln\frac{g^2\phi^2}{\Lambda^2}\times$$
\begin{equation}\label{eq5}
  \times\exp\left(i\int d^2x\left(\frac{1}{2}(\partial_{\mu}\phi)^2-\frac{m^2}{2}\phi^2-\frac{Ng^2\phi^2}{4\pi}\left(\ln\frac{g^2\phi^2}{\Lambda^2}-1\right)\right)\right),
\end{equation}
where $\Lambda$ is the ultraviolet cutoff. 

We want to obtain the chiral condensate in the framework of one-loop approximation. Therefore we calculate Eq.~(\ref{eq5}) using the method of the stationary phase. A minimum of the effective action of the system is reached if the effective potential and kinetic energy are minimal on its own:
\begin{equation}\label{eq6}
\partial_{\mu}\phi=0 \text{ and } U_{\text{eff}}(\phi)=\min.
\end{equation}
Let the constant scalar field $\phi_m$ satisfies condition (\ref{eq6}). The factor in front of the exponent in Eq.~(\ref{eq5}) is fixed at the point $\phi=\phi_m$ and we obtain
\begin{equation}\label{eq7}
  \left\langle0|g\bar{\psi}^a\psi^a|0\right\rangle=\frac{Ng^2\phi_m}{2\pi}\ln\frac{g^2\phi_m^2} {\Lambda^2}.
\end{equation}
The effective potential and the chiral condensate require renormalization. We renormalize effective potential following Coleman and Weinberg \cite{CW} and Gross and Neveu \cite{GN} by demanding that
\begin{equation}\label{eq8}
  \left.\frac{d^2U_{\text{eff}}}{d\phi_m^2}\right|_{\phi_m=M^2}=m_R^2.
\end{equation}
Then the renormalized chiral condensate is written as
\begin{equation}\label{eq9}
  \left\langle0|g\bar{\psi}^a\psi^a|0\right\rangle_R=\frac{Ng^2\phi_R}{2\pi}\ln\frac{\phi_R^2}{M^2}.
\end{equation}
$\phi_R$ is determined by mean the renormalized effective potential $U^R_{\text{eff}}$, which is written as
\begin{equation}\label{eq10}
U^R_{\text{eff}}=\frac{1}{2}m^2_R\phi^2_R+\frac{g^2N}{4\pi}\phi^2_R\left(\ln\frac{\phi^2_R}{M^2}-3\right),
\end{equation}
Here $\phi^2_R$ and $M^2$ are dimensionless.

The effective potential has the minimum, if
\begin{equation}\label{eq11}
\phi^2_m=M^2\exp 2\left(1-\frac{\pi m^2_R}{g^2N}\right).
\end{equation}
Note that this result coincides with that of refs.~\cite{GN}, \cite{KR1}.

The renormalized chiral condensate is written as
\begin{equation}\label{eq12}
\left\langle0|g\bar{\psi}^a\psi^a|0\right\rangle_R=\frac{Ng^2\phi_m}{2\pi}\ln\frac{\phi_m^2}{M^2}=\frac{N}{\pi}(g^2-g^2_{\text{cr}})\phi_m,
\end{equation}
where $g^2_{\text{cr}}=\frac{m^2_R\pi}{N}$. 

One can see the chiral condensate equals to zero, if $\phi^2_m=0$ or $\phi^2_m=M^2_R$, i.e. when $g^2=0$ or $g^2=g^2_{\text{cr}}$. In the latter case the discrete symmetry (\ref{eq2}) is not restored.

To find out with which physical phenomenon corresponds the second zero of the chiral condensate we expand the effective potential $U^R_{\text{eff}}(\phi)$ at the point $\phi^2_m=M^2$. We present $\phi^2_R(x)=M^2+\varphi^2$, where $\varphi^2$ describes small fluctuations ($\varphi^2/M^2\ll 1$) around the point $M^2$, and restrict ourselves to terms $\varphi^4$, then we get
$$U^R_{\text{eff}}(\phi^2)-U^R_{\text{eff}}(M^2)=\Delta U^R_{\text{eff}},$$
\begin{equation}\label{eq14}
\Delta U^R_{\text{eff}}=\left(m^2_R-\frac{g^2N}{\pi}\right)\frac{\varphi^2}{2}+\frac{g^2N}{2\pi M^2}\frac{\varphi^2}{4}.
\end{equation}
The mass term in $\Delta U^R_{\text{eff}}$ is $\mu^2=m^2_R-\frac{g^2N}{\pi}=\left(g^2_{\text{cr}}-g^2\right)\frac{N}{\pi}$ changes sign as a function of $g^2$. Really, when $g^2<g_{\text{cr}}^2$, $\mu^2>0$, and scalar field has a vacuum expectation value equals to zero $\langle\varphi^2\rangle=0$. If $g^2>g_{\text{cr}}^2$, $\mu^2<0$, we get $\langle\varphi^2\rangle=-\mu^2/\lambda^2$, where $\lambda^2=\frac{g^2N}{2\pi M^2}$. At the point $g^2=g_{\text{cr}}^2$ we have $\mu^2=0$, and $\phi^2_R=M^2$ (see Eq.~(\ref{eq11})). The chiral condensate also is equal to zero. This qualitative difference shows that the phase transition exists (\cite{GL}, \cite{W}, \cite{KR3}) and the chiral condensate is an order parameter.

Continuing our investigation of the model we want to take into account a small-interacting scalar field by adding term $\lambda\phi^4/24$, then the renormalized effective potential $U^R_{\text{eff}}$ in one-loop approximation is
\begin{equation}\label{eq15}
U^R_{\text{eff}}=\frac{m^2_R\phi^2_R}{2}+\frac{g^2N}{4\pi}\phi^2_R\left(\ln\frac{\phi^2_R}{M^2}-3\right)+\frac{\lambda\phi^4}{24}-\frac{\lambda M^2\phi^2_R}{4}-\frac{\lambda\phi^2_R}{16\pi}\left(\ln\frac{m^2_R+\frac{\lambda}{2}\phi^2_R}{m^2_R+\frac{\lambda}{2}M^2}-3\right).
\end{equation}
Here $U^R_{\text{eff}}$ satisfies demanding that
$$\left.\frac{d^2U^R_{\text{eff}}}{d\phi_R^2}\right|_{\phi^2_R=M^2}=m^2_R.$$
To study the $U^R_{\text{eff}}(\phi)$ analytically in the vicinity of $\phi^2_R=M^2$ we consider $(m^2_R-\frac{\lambda}{2}M^2)/m^2_R\ll 1$ and $\phi^2=M^2+\varphi^2$, $\varphi^2/M^2\ll 1$, then we get
$$\Delta U^R_{\text{eff}}(\varphi)=\frac{\mu\varphi^2}{2}+\frac{\lambda_0^2\varphi^4}{4},$$
where
\begin{equation}\label{eq16}
\mu^2=m^2_R-\frac{\lambda M^2}{3}-\frac{g^2N}{\pi}+\frac{5\lambda}{16\pi},\;\lambda_0^2=\frac{\lambda}{6}+\frac{g^2N}{2\pi M^2}-\frac{\lambda}{8\pi M^2}
\end{equation}
and $\langle \varphi^2\rangle=-\mu^2/\lambda^2_0$. This tells us, that the qualitative behavior of the theory does not change, if $\lambda/m^2_R\ll 1$.

\section{Conclusions}

In this paper we analyzed the zero points of the chiral condensate in model consisting of a scalar and a fermion fields with the Yukawa interaction in the framework of the one-loop approximation. The model is invariant under the discrete transformation. In our pervious paper \cite{KR2} we show that the chiral condensate vanishes at two different coupling constants. At the first of them there is a phase transition, which describes dynamical breaking of the discrete symmetry. In our paper we showed that the second zero of the chiral condensate describes the phase transition as a function of the coupling constant, which distinguishes the positive effective scalar mass from the negative one. The result is obtained analytically. We believe that this discussion can help to analyze similar phenomena in more realistic models.

\section*{Acknowlegdements}

We are grateful to O.V. Kancheli for useful discussions.

\end{document}